\documentclass[11pt,english]{extarticle}
\usepackage[latin9]{inputenc}
\usepackage{color}
\usepackage{array}
\usepackage{multirow}
\usepackage{amsmath}
\usepackage{amssymb}
\usepackage{stackrel}
\usepackage{graphicx}

\makeatletter

\providecommand{\tabularnewline}{\\}

\usepackage[center]{caption}
\usepackage[font=footnotesize]{caption}
\usepackage{flushend}

\usepackage{tikz}
\usetikzlibrary{mindmap}  

\usepackage{xcolor}
\usepackage[export]{adjustbox}

\AtBeginDocument{}
\usepackage{colortbl}
\definecolor{lightgray}{gray}{0.8}

\usepackage{setspace}
\usepackage[width=15cm]{geometry}

\usepackage{titling}
\makeatother

\usepackage{babel}
\begin{document}
\title{Autoencoder for Position-Assisted Beam Prediction in mmWave ISAC Systems }
\author{Ahmad A. Aziz El-Banna$^{1,}$$^{2}$, and Octavia A. Dobre$^{1}$
\\
\textit{ }$^{1}$Faculty of Engineering and Applied Science, Memorial
University, \\
 St. John\textquoteright s, Canada\\
$^{2}$Faculty of Engineering at Shoubra, Benha University, Egypt\\
E-mail: aaelbanna@mun.ca; odobre@mun.ca\thanks{This work was supported in part by Natural Sciences and Engineering
Research Council of Canada (NSERC), Discovery program RGPIN-2019-04123
and Canada Research Chair program CRC-2022-00187.}\vspace{-2mm}
}
\maketitle
\begin{abstract}
{\normalsize{}Integrated sensing and communication and millimeter
wave (mmWave) have emerged as pivotal technologies for 6G networks.
However, the narrow nature of mmWave beams requires precise alignments
that typically necessitate large training overhead. This overhead
can be reduced by incorporating the position information with beam
adjustments. This letter proposes a lightweight autorencoder (LAE)
model that addresses the position-assisted beam prediction problem
while significantly reducing computational complexity compared to
the conventional baseline method, i.e., deep fully connected neural
network. The proposed LAE is designed as a three-layer }\textit{\normalsize{}undercomplete}{\normalsize{}
network to exploit its dimensionality reduction capabilities and thereby
mitigate the computational requirements of the trained model. Simulation
results show that the proposed model achieves a similar beam prediction
accuracy to the baseline with an 83\% complexity reduction.}{\normalsize\par}
\end{abstract}
Index Terms--Integrated sensing and communication (ISAC), millimeter
wave (mmWave), autoencoder, beam prediction.

\section{\label{sec:Introduction}Introduction}

\def\figurename{Fig.}
\def\tablename{TABLE}

Integrated sensing and communication (ISAC), a promising technology
to be included in the 3GPP Release 19, is poised to be a key enabler
for 6G systems {[}\ref{R_ISAC6G}{]}--{[}\ref{R_10KML}{]}. By jointly
optimizing the spectrum and hardware resources, ISAC systems enable
a realistic co-design of communication and sensing tasks {[}\ref{R_ISAC70}{]}.
Moreover, the millimeter-wave (mmWave) technology, offering significant
improvements in spectral efficiency and spatial multiplexing capabilities,
is essential for current and future wireless systems {[}\ref{R_mmwavSrvy}{]}. 

Specific to the mmWave transmission is the narrow nature of the beams,
which necessitates precise alignments in order to provide adequate
coverage for users {[}\ref{R_BmPrdSmnl}{]}. This technical limitation
emphasizes the importance of robust beam prediction techniques in
mmWave-based communication systems. Moreover, the substantial training
overhead required for in-band channel estimation, particularly in
highly mobile scenarios, poses a significant challenge for beam tracking.
Nevertheless, given the imperative reliance of mmWave systems on line-of-sight
(LoS) links, exploiting user position information, e.g., global positioning
system (GPS), has emerged as a promising approach to reduce training
overhead and facilitate efficient beam adjustments {[}\ref{R_mmwavSrvy}{]},
{[}\ref{R_BmPrdSmnl}{]}.

Furthermore, in recent years, application of machine learning (ML)
has been proven to provide an effective solution to numerous communication
problems by enabling accurate modeling with reduced complexity compared
to traditional statistical models {[}\ref{R_10KML}{]}, {[}\ref{R_apprchs}{]},
{[}\ref{R_DeepSenseDst}{]}. The capabilities of ML techniques, in
particular neural networks (NNs), in identifying the optimal beam
for a sensing-aided mmWave system are investigated in {[}\ref{R_BmPrdSmnl}{]}.
While feed-forward NNs can effectively learn the complex mapping between
the user position and the optimal beam, its computational complexity
and memory footprint hinder its deployment on resource-constrained
equipment. This highlights the need for more computationally efficient
techniques, as real-time processing demands on modern devices continue
to grow.

To fully leverage the capabilities of ML techniques for beam prediction,
in this letter we propose a new lightweight autoencoder (LAE) architecture
that significantly reduces the computational cost of state-of-the-art
models, achieving approximately an 83\% reduction in multiplications,
additions, and model parameters. We design a computationally efficient
LAE model, comprising only three hidden layers, and systematically
select its configuration parameters to ensure robust generalization
across multiple datasets that represent diverse environments of mmWave
ISAC systems. The key contributions of this work are: (i) the development
of a structured, low-complexity LAE tailored for beam prediction on
resource-constrained devices; (ii) a comprehensive complexity analysis
demonstrating substantial reductions in computational requirements;
and (iii) empirical validation showing that the proposed model achieves
prediction accuracy comparable to NNs across a wide range of real-world
scenarios.

\begin{figure}
\begin{centering}
\includegraphics[scale=0.07]{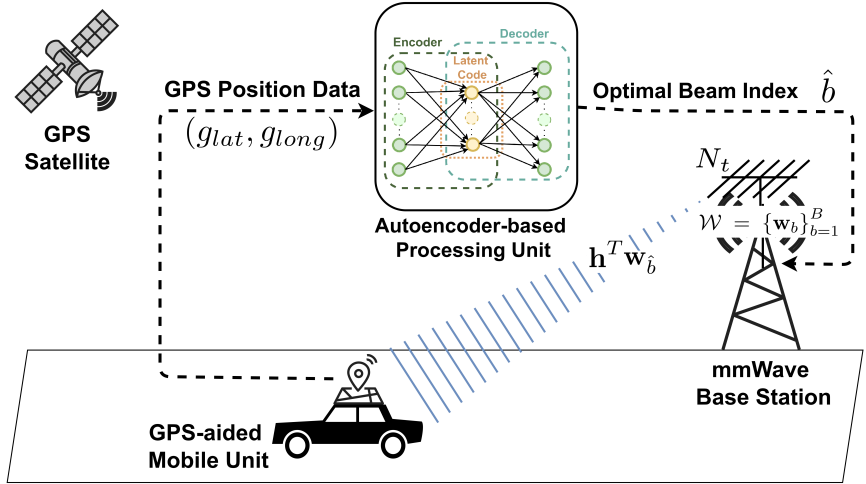}
\par\end{centering}
\centering{}\caption{\label{fig:SystemModel}A system model integrating the autoencoder
beam prediction model with mmWave ISAC system.}
\end{figure}
\vspace{-2mm}

\section{\label{sec:SysMdl}System Model and Problem Definition}

\subsection{System Model }

We consider a mmWave ISAC system that utilizes an autoencoder for
position-assisted beam alignment, as depicted in Fig. \ref{fig:SystemModel}.
A single-antenna mmWave mobile unit (MU), equipped with GPS, transmits
its location information to an autoencoder-based processing unit which
aids an $N_{t}$-antenna stationary mmWave base station (BS) in selecting
one of the $B$ beamforming vectors, $\mathbf{w}_{b}{\color{black}\in\mathbb{C}^{N_{t}\times1}}$,
from a codebook $\mathcal{W}=\left\{ \mathbf{w}_{b}\right\} _{b=1}^{B}$,
in order to transmit a complex symbol $x$. The received downlink
signal can be expressed as $y=\boldsymbol{\mathbf{h}}^{T}\mathbf{w}_{b}x+n$,
where $\mathbf{h}\in\mathbb{C}^{N_{t}\times1}$ represents the channel
vector between the mmWave BS antennas and the mmWave MU antenna, and
$n\sim\mathcal{N_{\mathbb{C}}}\left(0,\sigma^{2}\right)$ is the complex-valued
Gaussian distributed noise with zero mean and variance $\sigma^{2}$.
\vspace{-2mm}

\subsection{Problem Definition }

Considering the received signal power as the evaluation metric, the
optimal beam maximizes the expected received power $P=\mathbb{E}\left[\left|y\right|^{2}\right]$
at the designated user. Consequently, the ideal beamformer $\mathbf{w}^{\star}$
is selected to receive the maximum power $P$ at the BS, and the problem
can be formulated as $\mathbf{w}^{\star}=\underset{\mathbf{w}\in\mathcal{W}}{\arg\max}\left|\boldsymbol{\mathbf{h}}^{T}\mathbf{w}\right|^{2}${[}\ref{R_BmPrdSmnl}{]}.
The typical challenges associated with acquiring the channel information
$\mathbf{h}$, and the reliance of mmWave systems on narrow, directional
beams and LoS paths, motivate using real-time position information
as an alternative to the direct channel estimation for the beam selection,
i.e., bypassing the explicit estimation of $\boldsymbol{\mathbf{h}}^{T}$
{[}\ref{R_BmPrdSmnl}{]}. The problem now shifts to approximate $\mathbf{w}^{\star}$
by estimating $\hat{\mathbf{w}}$ that maximizes $\mathbb{P\left(\mathit{{\color{black}{\color{blue}{\color{black}\hat{\mathbf{w}}=\mathbf{w}^{\star}}}|\mathbf{p}}}\right)}$,
where $\mathbf{p}=\left(g_{lat},g_{long}\right)$ denotes the 2-D
position vector, with $g_{lat}$ and $g_{long}$ as the latitude and
longitude GPS coordinates, respectively.

By collecting proper observations of the position and corresponding
optimal beam pairs, ML techniques can be effectively utilized to learn
the mapping between position and beam. Using a captured $Q$-labeled
dataset, i.e., $\mathcal{Q}=\left\{ \left(\mathbf{p}_{q},\mathbf{w}_{q}^{\star}\right):q=1,\cdots,Q\right\} $,
an ML algorithm can estimate the class probability vector $\mathcal{C}\in\left\{ c_{1},\cdots,c_{B}\right\} $,
with $c_{b}=\mathbb{P}\left(\mathbf{w}_{b}=\mathbf{w}^{\star}\right)$
and $b\in\left\{ 1,\cdots,B\right\} $, in order to choose the beamformer
with the highest probability, i.e., $\hat{\mathbf{w}}=\mathbf{w}_{\hat{b}},\hat{b}=\underset{b\in\left\{ 1,\cdots,B\right\} }{\arg\max}c_{b}$.\vspace{-2mm}

\section{\label{sec:PrpsdAECnfig}Proposed Autoencoder (AE) Architecture}

\begin{figure}
\begin{centering}
\includegraphics[scale=0.09]{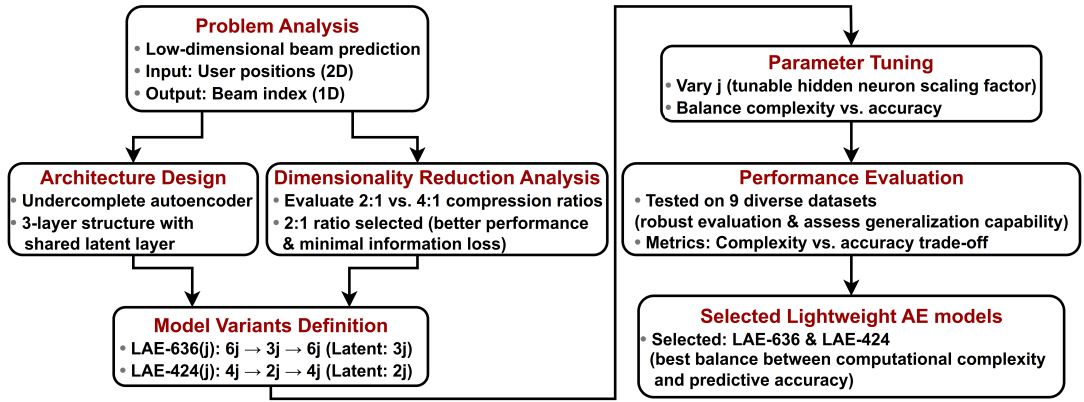}
\par\end{centering}
\centering{}\caption{\label{fig:LAESlctnWrkFlw}LAE design and selection workflow.}
\end{figure}

An AE is an NN architecture composed of two primary components: an
encoder and a decoder {[}\ref{R_AESmnl}{]}. Typically, the encoder
compresses the input data into a compact latent code, while the decoder
aims to reconstruct the original data from this latent code. The training
objective is to minimize the reconstruction loss, which measures the
discrepancy between the reconstructed and original data.

In this work, we utilize the dimensionality reduction properties of
an \textit{undercomplete} AE structure, wherein the code space has
fewer dimensions than the input space, to learn a compact representation
of the data that captures the most pertinent features, leading to
a low-complexity model. Moreover, our proposed model is a lightweight
AE named LAE, composed of three layers, with the middle (latent) layer
shared between the encoder and decoder, as depicted in Fig. \ref{fig:SystemModel}. 

Given that the input features represent user positions (two correlated
values) and the target is a single beam index, the beam prediction
problem is inherently low-dimensional. To ensure minimal information
loss during compression, a 2:1 ratio is chosen between the encoder/decoder
layers and the latent layer, after evaluating both 2:1 and 4:1 ratios
and determining that 2:1 provides better performance. To further balance
complexity and accuracy, we conducted systematic experimentation and
analysis across multiple autoencoder configurations, ultimately selecting
two lightweight models, namely LAE-636 and LAE-424, that offer favorable
trade-offs between computational complexity and predictive accuracy.

Specifically, LAE-636($j$) is a 3-layers AE that utilizes a 2:1 compression
ratio and a latent dimension of $3j$, resulting in an architecture
with $6j$, $3j$, and $6j$ hidden neurons in the first, latent,
and last layers, respectively, where $j$ is a predetermined number
of hidden neurons. LAE-424($j$) follows a similar structure but with
a latent dimension of $2j$. As examples, LAE-636(10) is a basic AE
with an initial layer consisting of 60 neurons, a latent space dimension
of 30, and a final layer comprising 60 neurons, while LAE-424(10)
is an AE with a 40-20-40 structure.

Finally, we trained and tested the models on nine datasets collected
under diverse conditions, including variations in time of day, weather,
and data volume. This demonstrates the robustness and generalization
capability of the proposed LAEs while reducing the risk of overfitting
to any specific scenario. The LAE design and selection workflow is
shown in Fig. \ref{fig:LAESlctnWrkFlw}. 

\vspace{-2mm}

\section{Dataset Analysis and Best LAE Configurations Selection}

\subsection{\label{subsec:Dataset}Dataset Analysis}

In this work, we utilize the dataset from {[}\ref{R_BmPrdSmnl}{]},
which is part of the comprehensive DeepSense dataset for 6G learning
research {[}\ref{R_DeepSenseDst}{]}. All nine scenarios outlined
in {[}\ref{R_BmPrdSmnl}{]} are employed to conduct a thorough evaluation
of the performance of the proposed LAE models in comparison to the
baseline in {[}\ref{R_BmPrdSmnl}{]}, namely FCN3(256).\footnote{We select FCN3(256), a three-layer NN with 256 hidden neurons, as
a baseline due to its superior performance within the evaluated ML
models {[}\ref{R_BmPrdSmnl}{]}.} Two units were used to collect the dataset, emulating the mmWave
ISAC system shown in Fig. \ref{fig:SystemModel}. The first unit,
a stationary mmWave BS equipped with a 16-element 60 GHz-band phased
array, outputs a 64-element vector, along with the received power
at each beam.\footnote{The first unit includes additional hardware like cameras and radar,
but this work focuses on position-to-beam mapping using beam power
data and GPS coordinates. The BS uses analog beamforming with a 16-element
phased array, selecting one beam from a 64-beam codebook per user.
The dataset is single-user, as per the testbed design. Details on
the hardware and system setup are in {[}\ref{R_tstbd}{]}. Extending
this framework to multi-user beam prediction is a promising future
direction requiring additional data and system-level considerations.
Moreover, future work could explore sensor fusion approaches, such
as integrating GPS with inertial, radar, or vision sensors, to improve
localization robustness and beam prediction accuracy under varying
localization precision and environmental conditions {[}\ref{R_DeepSenseDst}{]},
{[}\ref{R_OmniCNN}{]}.} The second unit, a mobile vehicle, represents the MU and is equipped
with a 60 GHz omni-directional mmWave transmitter and a GPS receiver
capable of outputting 10 location measurements per second and supports
real-time kinematic (RTK) for centimeter-level accuracy. However,
since the exact GPS mode used during data collection is not specified,
we conservatively consider a typical differential GPS accuracy ($\sim$2.5\,m)
as the upper bound during model training. However, it worth mentioning
that the proposed model is trained on real-world data collected under
these conditions, allowing it to implicitly learn and adapt to the
inherent imperfections in positioning accuracy. Finally, the dataset
was collected from multiple locations across the Arizona State University
campus, capturing a diverse range of real-world scenarios. 

A summary of the features and the number of data samples collected
for each scenario is presented in Table \ref{tab:ScnrChrs}. As can
be observed in this table, the nine scenarios encompass a diverse
range of operating conditions, with each scenario comprising hundreds
to thousands of data samples. Scenarios 5, 6, and 7 present the most
challenging conditions for training ML models. The dataset for scenarios
6 and 7 contains a limited number of captured samples, which can potentially
constrain the learning capabilities of ML algorithms, particularly
those whose performance is significantly influenced by the size of
the training dataset, such as NNs. Further, the data collected for
scenario 5 was acquired under rainy weather conditions, which can
degrade the mmWave signal propagation due to increased absorption
and scattering losses. This can lead to noisy data, making it more
difficult for the ML algorithms to learn the underlying position-beam
mapping function. 

Finally, it is important to note that although no dataset can fully
represent all possible deployment scenarios, by training and testing
the model across this diverse set of conditions, we ensure that the
learned representations are not tailored to any single data distribution.
Instead, the model learns to capture underlying spatial and propagation
patterns that are common across environments, resulting in a generic
and deployment-ready solution.

\begin{table}
{\scriptsize{}\caption{\textcolor{red}{\label{tab:ScnrChrs}}Features of the nine dataset
scenarios for street-level mmWave communication {[}\ref{R_DeepSenseDst}{]},
{[}\ref{R_tstbd}{]}.}
}{\scriptsize\par}
\begin{centering}
\vspace{0mm}
\begin{tabular}{|c|c|c|}
\hline 
{\scriptsize{}Category Description} & {\scriptsize{}Scenarios} & {\scriptsize{}\# Samples}\tabularnewline
\hline 
\hline 
\multirow{6}{*}{{\scriptsize{}Day-time Scenarios}} & {\scriptsize{}1} & {\scriptsize{}2,411}\tabularnewline
\cline{2-3} \cline{3-3} 
 & {\scriptsize{}3} & {\scriptsize{}1,487}\tabularnewline
\cline{2-3} \cline{3-3} 
 & {\scriptsize{}6} & {\scriptsize{}915}\tabularnewline
\cline{2-3} \cline{3-3} 
 & {\scriptsize{}7} & {\scriptsize{}854}\tabularnewline
\cline{2-3} \cline{3-3} 
 & {\scriptsize{}8} & {\scriptsize{}4,043}\tabularnewline
\cline{2-3} \cline{3-3} 
 & {\scriptsize{}9} & {\scriptsize{}5,964}\tabularnewline
\hline 
\multirow{2}{*}{{\scriptsize{}Night-time Scenarios}} & {\scriptsize{}2} & {\scriptsize{}2,974}\tabularnewline
\cline{2-3} \cline{3-3} 
 & {\scriptsize{}4} & {\scriptsize{}1,867}\tabularnewline
\hline 
{\scriptsize{}Night-time Scenario, Rainy Weather Condition} & {\scriptsize{}5} & {\scriptsize{}2,300}\tabularnewline
\hline 
\end{tabular}
\par\end{centering}
\centering{}\vspace{-4mm}
\end{table}
\vspace{-3mm}

\subsection{\label{subsec:Optimum-LAE-Configurations}Best LAE Configurations
Selection}

\begin{figure*}
\begin{centering}
\includegraphics[scale=0.1]{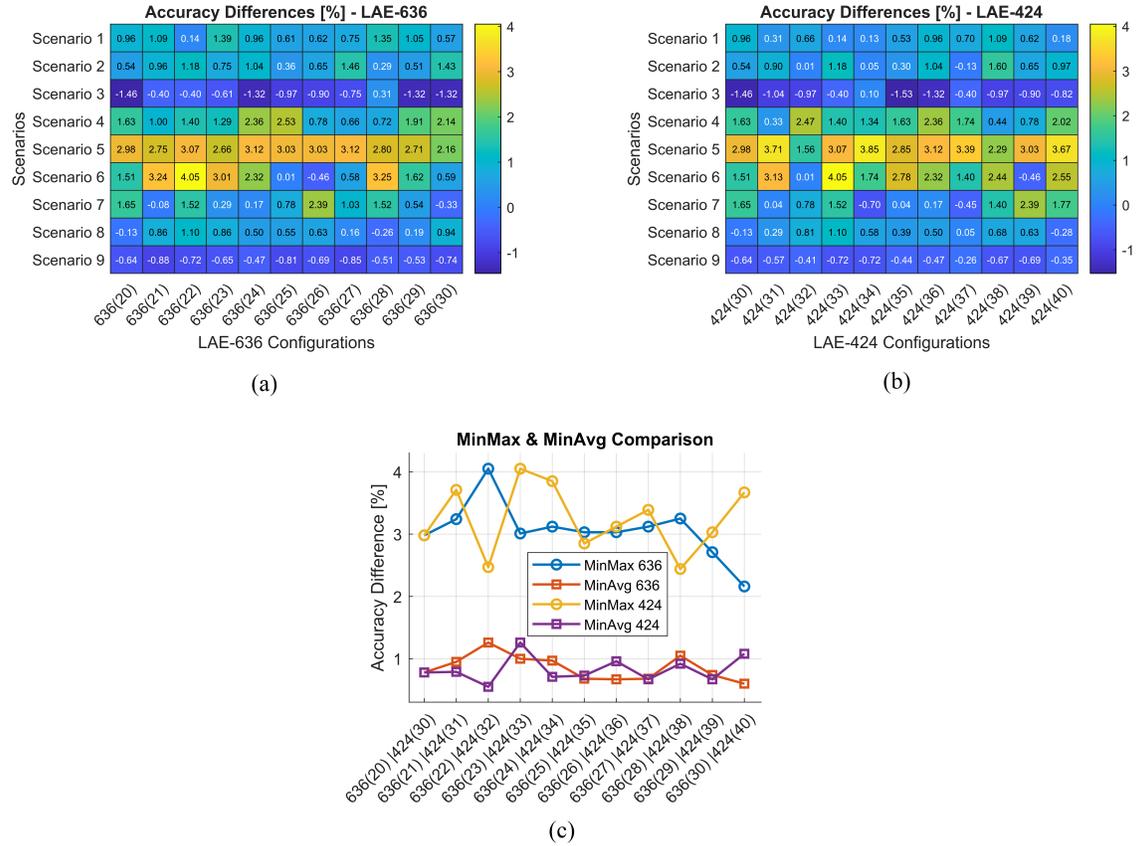}
\par\end{centering}
\centering{}\caption{\label{fig:htmp_lae636424}Accuracy differences {[}\%{]} between the
baseline, i.e., FCN3(256), and configuration (1), i.e., LAE-636, (a)
and configuration (2), i.e., LAE-424, (b). \textit{MinMaxDiff} and
\textit{MinAvgDiff} values for both LAE-636 and LAE-424 configurations
(c).}
\end{figure*}

\subsection*{Configuration (1): LAE-636}

In this analysis, we examine the LAE-636 structure with a range of
hidden neuron values $j=\{20,21,\ldots,30\}$\footnote{We tested other neuron structures outside the range of 20-30 for AE-636
and 30-40 for AE-424. However, these structures either performed worse
or similarly to those within the specified ranges, while also being
more complex. Therefore, we focused our analysis on the 20-30 range
for AE-636 and 30-40 range for AE-424.} and compare the achieved performance, i.e., top-1 beam prediction
accuracy, of each structure to the baseline FCN3(256). Figure \ref{fig:htmp_lae636424}
(a) presents a heatmap illustrating the accuracy differences between
the baseline and the tested LAE-636 structures for each of the nine
scenarios described above.

In addition, we employed the \textit{minimum-maximum difference (MinMaxDiff)}
evaluation metric to determine the largest accuracy difference between
the baseline and LAE structures for each scenario. Besides, the \textit{minimum-average
difference (MinAvgDiff)} metric was used to further differentiate
the models by calculating the average of the accuracy differences
between the LAE structures and the baseline across all the entire
nine scenarios. The former metric identifies the model with the largest
accuracy drop, while the latter provides an overall assessment of
the model's performance relative to the baseline, taking into account
both gains and losses across all testing scenarios. The \textit{MinAvgDiff}
metric also considers the individual accuracy gains of the LAE structures
in specific scenarios. For instance, LAE-636(30) exhibits accuracy
gains of 1.32\%, 0.33\%, and 0.74\% over the baseline in scenarios
3, 7, and 9, respectively, demonstrating performance advantages not
evident in the \textit{MinMaxDiff} metric. 

Finally, LAE-636(30) is selected for further analysis due to its optimal
performance in terms of both \textit{MinMaxDiff} and \textit{MinAvgDiff},
as shown in Fig. \ref{fig:htmp_lae636424} (c).\vspace{-2mm}

\subsection*{Configuration (2): LAE-424}

Similarly, the performance of the LAE-424 structure was assessed for
a range of hidden neurons $j=\{30,31,\ldots,40\}$. The achieved top-1
beam prediction accuracy of each structure was compared to the baseline
FCN3(256) model. Figure \ref{fig:htmp_lae636424} (b) provides a detailed
comparison of the accuracy differences between the examined LAE-424
structures and the baseline for each scenario. The LAE-424(38) and
LAE-424(32) structures are selected as they exhibit the lowest \textit{MinMaxDiff}
and \textit{MinAvgDiff}, respectively, as shown in Fig. \ref{fig:htmp_lae636424}
(c).

In the analysis of Fig. \ref{fig:htmp_lae636424}, hyperparameter
tuning is performed to optimize the LAE models. The following hyperparameters
are examined: batch size (BZ) $=\left\{ 4,8,16,32,64,128\right\} $,
learning rate (LR) $=\left\{ 0.0001,0.001,0.01\right\} $, and activation
function (AF) $=\left\{ Sigmoid,ReLU\right\} $. The optimal hyperparameter
values are found to be $BZ=16$, $LR=0.01$, and $AF=ReLU$.\footnote{Notably, the main framework code and the datasets employed in this
study are publicly accessible online {[}\ref{R_BmPrdSmnl}{]}, {[}\ref{R_tstbd}{]}.}\vspace{-2mm}

\section{Results and Discussion}

\subsection{Numerical Results}

In this subsection, we analyze the top-1 and top-k beam prediction
accuracies to evaluate the performance of the proposed LAE models
in comparison to the baseline. The large number of employed beams
(64) transforms the beam prediction problem into a challenging multi-label
classification task. As such, to provide a more lenient measure of
performance and assess model robustness, we include both top-1 and
top-k\footnote{Regardless of the primary objective of identifying the most effective
beam, i.e., the top-1 predicted beam, the selection of subsequent
beams can still contribute to covering the designated user, although
they are not the optimal choice.} beam prediction accuracy as performance indicators. In addition to
the analysis of beam prediction accuracy, we analyze the power loss
between the predicted and actual beams, and the effect of changing
the codebook size on the model performance.

\begin{figure}
\begin{centering}
\includegraphics[scale=0.13]{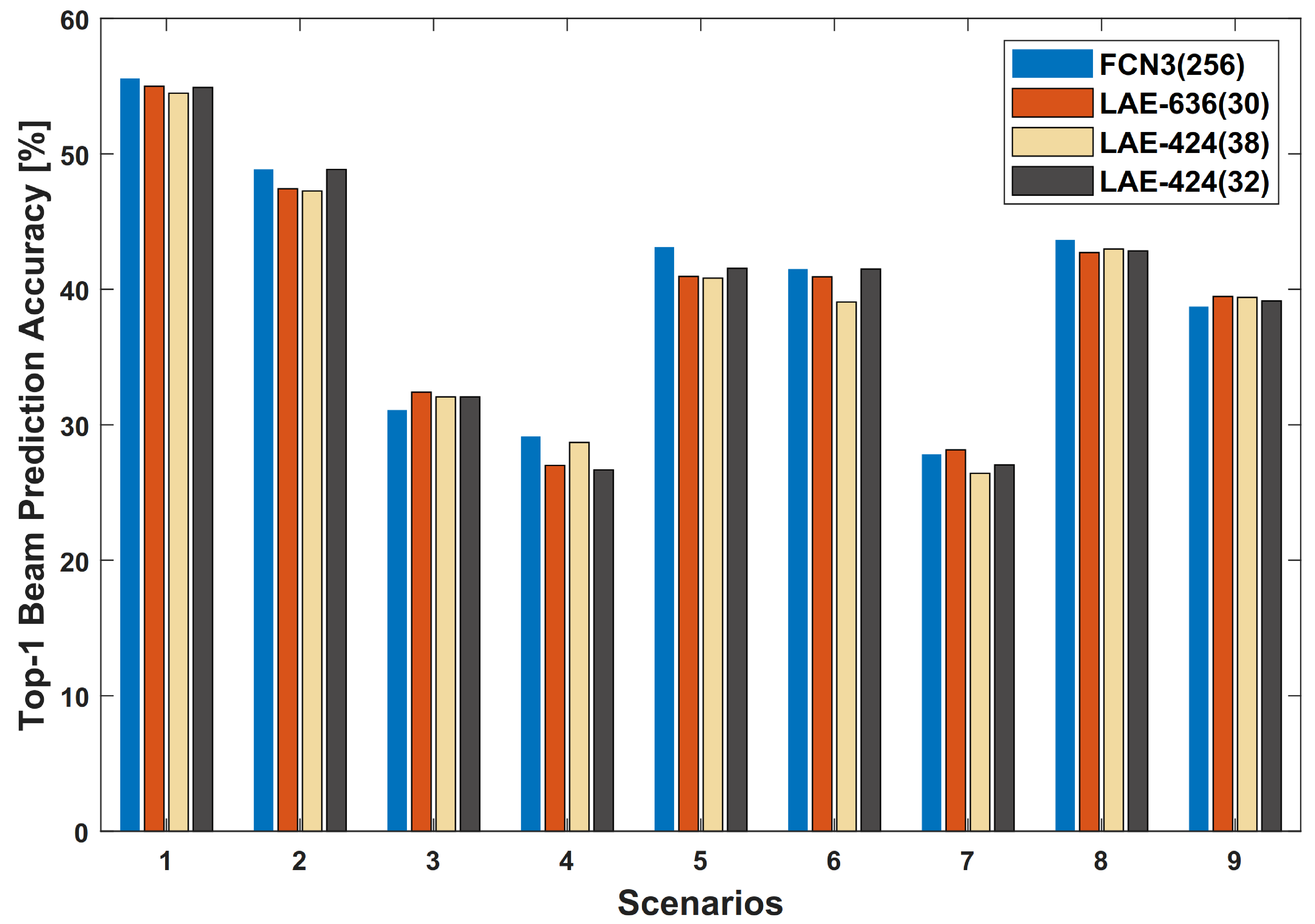}
\par\end{centering}
\caption{\label{fig:Top-1-acc}Top-1 beam prediction accuracy for the LAE and
baseline models.}
\vspace{-2mm}
\end{figure}

Utilizing the aforementioned dataset, we optimize the LAE structures
to achieve performance comparable to that of the baseline across the
entire range of the real-world scenarios. Figure \ref{fig:Top-1-acc}
shows the top-1 beam prediction accuracy of the pre-selected LAE models,
i.e., LAE-636(30), LAE-424(38), and LAE-424(32), compared to the FCN3(256)
baseline, for all scenarios. As depicted in Fig. \ref{fig:Top-1-acc},
the selected LAE models perform similarly to the baseline across all
nine scenarios. Moreover, Fig. \ref{fig:htmp_lae636424} demonstrates
that the three LAE models achieve a minimal accuracy gap of up to
2.47\% relative to the baseline. 

\begin{figure*}
\begin{centering}
\includegraphics[scale=0.093]{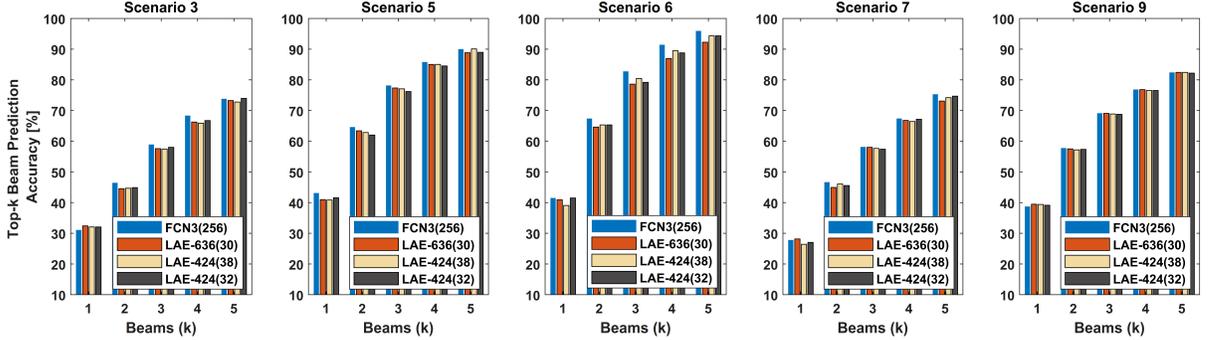}
\par\end{centering}
\caption{\label{fig:Top-5-acc}Top-k beam prediction accuracy for scenarios
3, 5, 6, 7 and 9.}
\end{figure*}

Additionally, Fig. \ref{fig:Top-5-acc} provides a comparative analysis
of the top-k beam accuracy for the different models, with a focus
on scenario 5 (rainy conditions), scenarios 3, 6, and 7 (small datasets),
and scenario 9 (large dataset). As depicted in Fig. \ref{fig:Top-5-acc},
the LAE and the baseline models exhibit similar generalization patterns,
emphasizing the comparable performance achieved by the proposed LAE
architecture. We select the LAE-424(32) model for the upcoming analysis
as it attains the closest performance to the baseline in terms of
both top-1 and top-k beam prediction accuracy.

\begin{table*}
\vspace{2.5mm}
{\scriptsize{}\caption{\textcolor{red}{\label{tab:Pwr}} \textcolor{black}{Power loss for
the propos}ed LAE-424(32) and \textcolor{black}{baseline models.}}
}{\scriptsize\par}

\vspace{-1mm}

\centering{}%
\begin{tabular}{|c|c|c|c|c|c|c|c|c|c|}
\hline 
{\tiny{}$L_{[dB]}$} & {\tiny{}Scenario 1} & {\tiny{}Scenario 2} & {\tiny{}Scenario 3} & {\tiny{}Scenario 4} & {\tiny{}Scenario 5} & {\tiny{}Scenario 6} & {\tiny{}Scenario 7} & {\tiny{}Scenario 8} & {\tiny{}Scenario 9}\tabularnewline
\hline 
\hline 
\textbf{\tiny{}FCN3(256)} & {\scriptsize{}0.27} & {\scriptsize{}0.43} & {\scriptsize{}2.05} & {\scriptsize{}2.46} & {\scriptsize{}1.04} & {\scriptsize{}0.36} & {\scriptsize{}1.56} & {\scriptsize{}1.22} & {\scriptsize{}2.63}\tabularnewline
\hline 
\textbf{\tiny{}LAE-424(32)} & {\scriptsize{}0.26} & {\scriptsize{}0.45} & {\scriptsize{}2.32} & {\scriptsize{}2.44} & {\scriptsize{}1.14} & {\scriptsize{}0.33} & {\scriptsize{}1.46} & {\scriptsize{}1.25} & {\scriptsize{}2.6}\tabularnewline
\hline 
\end{tabular}
\end{table*}

Further, a rigorous metric for assessing system performance is the
power loss between predicted and actual beam; the average power loss
$L$ can be expressed as $L_{[dB]}=10\log_{10}\left(\frac{1}{Q}\stackrel[q=1]{Q}{\sum}\frac{P_{\boldsymbol{\mathbf{w}}^{\star}}^{q}-P_{n}}{P_{\boldsymbol{\mathbf{\hat{\mathbf{w}}}}}^{q}-P_{n}}\right)$,
where $P_{\boldsymbol{\mathbf{w}}^{\star}}^{q}$ and $P_{\boldsymbol{\mathbf{\hat{\mathbf{w}}}}}^{q}$
are the ground truth beam and predicted beam powers, respectively,
and $P_{n}$ is the scenario noise power {[}\ref{R_BmPrdSmnl}{]}.
Table \ref{tab:Pwr} presents a comparative analysis of the power
loss for both the selected LAE-424(32) model and the baseline model.
As inferred from the table, the proposed LAE-424(32) model exhibits
comparable performance to the baseline across the entire range of
scenarios and it further outperforms the baseline in five scenarios,
i.e., scenarios 1, 4, 6, 7, and 9.

\begin{figure}
\begin{centering}
\includegraphics[scale=0.13]{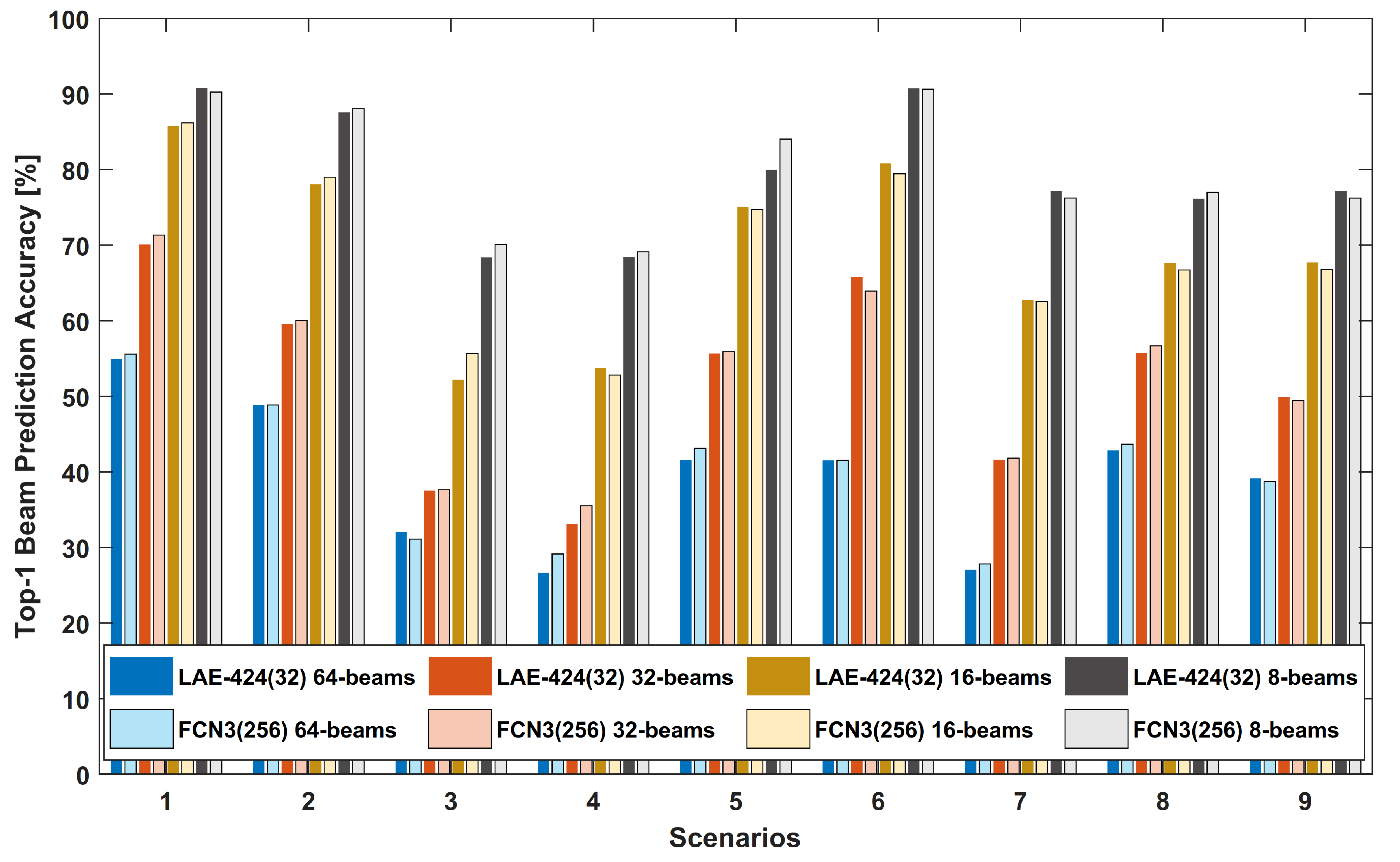}
\par\end{centering}
\caption{\label{fig:Bms}Top-1 beam prediction accuracy at different codebook
sizes (number of beams) for the proposed LAE and baseline models.}
\end{figure}

Finally, we evaluate the top-1 beam prediction accuracy of the proposed
LAE-424(32) and baseline models for different codebook sizes in Fig.
\ref{fig:Bms}. The figure shows that the proposed LAE-424(32) model
acquires similar performance to the baseline across different numbers
of employed beams. Notably, the proposed LAE-424(32) model exhibits
superior performance to the baseline for a greater number of scenarios
when smaller codebook sizes are employed.

\vspace{-3mm}

\subsection{Complexity Evaluation}

An NN-based model constructed from a set of fully-connected dense
layers $S$ requires a total number of parameters equal to $\theta_{D}={\color{black}\sum}_{i=1}^{S}\theta_{di}$,
with $\theta_{di}=m_{di}^{hd}\left(2m_{d(i-1)}^{hd}+1\right)$, where
$m_{di}^{hd}$ is the number of hidden neurons in layer $i$ and $m_{d(i-1)}^{hd}$
is the number of hidden neurons in its previous layer $i-1$ {[}\ref{R_hybrd}{]}.
Similarly, the number of required multiplications (MUL) or additions
(ADD) can be computed as ${\color{blue}{\color{black}\sum}_{{\color{black}i=1}}^{{\color{black}{\color{blue}{\color{black}S}}}}}m_{di}^{hd}m_{d(i-1)}^{hd}$
{[}\ref{R_hybrd}{]}.\footnote{Both the FCN3(256) and AE models employ a ReLU activation function,
which is assumed to require a single floating-point operation {[}\ref{R_hybrd}{]}.
Consequently, the computational cost associated with ReLU activations
is neglected in our calculations.}

Table \ref{tab:FinlCmprsn} presents the complexity reduction achieved
by the candidate LAE models compared to the baseline. As can be seen
from the table, the highest complexity reduction of over 83\% in terms
of the number of multiplications, additions, and memory requirements,
i.e., model parameters, is offered by the LAE-424(32) model. The remaining
LAE models also demonstrate significant complexity reduction, ranging
from approximately 70\% to 77\%. 

Eventually, the proposed LAE-424(32) model offers a significant complexity
reduction of over 83\% compared to the baseline, while maintaining
a high level of performance with less than 2.5\% loss in beam prediction
accuracy as evident from Table \ref{tab:FinlCmprsn}. Furthermore,
LAE-424(32) has an average accuracy degradation of only 0.55\% compared
to the baseline across all tested scenarios, as indicated in Table
\ref{tab:FinlCmprsn}, which demonstrates the model's ability to generalize
to different environmental conditions, highlighting its robustness
and real-world applicability.

In summary, this work introduced an optimized LAE architecture for
beam prediction in mmWave ISAC systems, with the LAE-424(32) model
emerging as a particularly effective configuration. Through systematic
structural tuning and parameter selection, the model achieves accuracy
on par with the baseline while delivering substantial reductions in
computational and memory demands, demonstrating the effectiveness
of architectural optimization in enabling lightweight and practical
deployment.

\begin{table}
\vspace{2mm}
\caption{\label{tab:FinlCmprsn}\textcolor{black}{Quantitative comparison in
terms of complexity and accuracy.}}

\begin{centering}
\begin{tabular}{|c|c|c|c|c|}
\hline 
{\tiny{}Criteria} & {\tiny{}Model} & \textbf{\tiny{}LAE-636(30)} & \textbf{\tiny{}LAE-424(38)} & \textbf{\tiny{}LAE-424(32)}\tabularnewline
\hline 
\hline 
\multirow{2}{*}{\textbf{\tiny{}Complexity Reduction}} & {\tiny{}MUL/ADD {[}\%{]}} & {\scriptsize{}70.07} & {\scriptsize{}77.61} & \textbf{\scriptsize{}83.22}\tabularnewline
\cline{2-5} \cline{3-5} \cline{4-5} \cline{5-5} 
 & {\tiny{}Parameters {[}\%{]}} & {\scriptsize{}69.90} & {\scriptsize{}77.43} & \textbf{\scriptsize{}83.05}\tabularnewline
\hline 
\multirow{2}{*}{\textbf{\tiny{}Accuracy Difference}} & {\tiny{}Maximum {[}\%{]}} & \textbf{\scriptsize{}2.16} & {\scriptsize{}2.44} & {\scriptsize{}2.47}\tabularnewline
\cline{2-5} \cline{3-5} \cline{4-5} \cline{5-5} 
 & {\tiny{}Average {[}\%{]}} & {\scriptsize{}0.60} & {\scriptsize{}0.92} & \textbf{\scriptsize{}0.55}\tabularnewline
\hline 
\end{tabular}
\par\end{centering}
\vspace{-2mm}
\end{table}
\vspace{-2mm}

\section{\label{sec:Concolsn}Conclusion}

In this letter, we proposed a lightweight autoencoder (LAE) model
for position-assisted beam prediction. The proposed LAE model, constructed
from an \textit{undercomplete} structure with only three layers, leverages
dimensionality reduction to efficiently learn the mapping between
position and beam. Evaluated on a comprehensive real-world dataset
encompassing nine distinct mmWave ISAC communication scenarios, the
proposed model demonstrates comparable performance to the baseline,
achieving an average beam prediction accuracy loss of only 0.55\%
while significantly reducing computational complexity by over 83\%.

\end{document}